\documentclass[prd,preprint,a4paper,superscriptaddress,nofootinbib,11pt]{revtex4}
\usepackage{amsmath,amsfonts,amssymb}
\usepackage{txfonts}
\usepackage[T1]{fontenc}
\renewcommand{\d}{\mathrm{d}}
\begin{document}


\begin{center}
{\bf  \Large Universal $\kappa$-Poincar\'e covariant differential calculus over $\kappa$-Minkowski space}
 
 \bigskip
\bigskip

Tajron Juri\'c, {\footnote{e-mail: tjuric@irb.hr}} Stjepan Meljanac, {\footnote{e-mail: meljanac@irb.hr}} \\  
Rudjer Bo\v{s}kovi\'c Institute, Bijeni\v cka  c.54, HR-10002 Zagreb,
Croatia \\[3mm]

Rina  \v{S}trajn {\footnote{e-mail: ri.strajn1@studenti.unica.it}},
\\
Dipartimento di Matematica e Informatica,
Universita di Cagliari,\\
 viale Merello 92, I-09123 Cagliari, Italy
and INFN, Sezione di Cagliari
 \\[3mm]

\end{center}
\setcounter{page}{1}
\bigskip

{
Unified graded differential algebra, generated by $\kappa$-Minkowski noncommutative (NC) coordinates, Lorentz generators and anticommuting one-forms, is constructed.
It is compatible with $\kappa$-Poincar\'e-Hopf algebra. For time- and space-like deformations, the super-Jacobi identities are not satisfied.
By introducing additional generator, interpreted as exterior derivative, we find a new unique algebra that satisfies all super-Jacobi identities.
It is universal and valid for all type of deformations (time-, space-, and light-like). For time-like deformations this algebra coincides with the one in \cite{sitarz}.
Different realizations of our algebra in terms of super-Heisenberg algebra are presented. For light-like deformations we get 4D bicovariant calculus, with $\kappa$-Poincar\'e-Hopf algebra
and present the corresponding twist, which is written in a new covariant way, using Poincar\'e generators only. In the time- and space-like case this twist leads to $\kappa$-Snyder space. 
Our results might lead to applications in NC quantum field theories (especially electrodynamics and gauge theories), quantum gravity models, and Planck scale physics.
}



\newpage
\section{Introduction}
One of the open problems in theoretical physics is the search for a fundamental theory describing physics at the Planck scale, i.e. the scale at which gravitational effects are equally strong as quantum ones. At the Planck scale the idea of size or distance in classical terms is no longer valid, because one has to take into account quantum uncertainty \cite{a1}. Such effects are naturally treated in the noncommutative (NC) spacetime approach and one needs to use the framework of noncommutative geometry (NCG). Noncommutative geometry, particularly NC spacetimes, have found many applications in physical theories, concerning NCQFT defined over NC spaces with twisted Poincare symmetry \cite{a2}, NC gauge theories \cite{a3} and NC gravity \cite{a4, a5, a6}. Deformation of spacetime and its symmetries \cite{a1}, \cite{a7} is considered as one of the approaches to quantization of the theory of gravity. 
  
	Research on the structure of spacetime at the Planck scale, where quantum gravity corrections have to be taken into account, is one of the most important challenges of fundamental physics. The Hopf algebra (quantum group) formalism is strongly connected with NC spacetimes. Quantum groups \cite{majidknjiga} are thought as generalizations of symmetry groups of the underlying spacetime. The quantum group formalism was also fruitfully used in quantizing gravity (in 2+1 dimensions) starting from the Chern-Simons formulation. In that case the relevant quantum group is the Lorentz double, which is the quantum double of the universal cover of the (2+1)-dimensional Lorentz group \cite{a8}. Recently there has been interest in the noncommutative version of Riemannian geometry \cite{a9, a10, a11}. One of the most fruitful approaches has been proposed by the J. Wess and collaborators \cite{a3, a4}, which combined with quantum group methods resulted in the formulation of twist-deformed Riemannian geometry together with the twisted version of Einstein equations \cite{a9}.
	
	Among the quantum symmetries, the $\kappa$-Poincar\'e symmetry \cite{a7, majid, z1994} is one of the most extensively studied together with the $\kappa$-deformed Minkowski space  emerging out of it through the cross product algebra construction \cite{majid}. The reason for this is two-fold. First, the quantum field theory with $\kappa$-Poincar\'e symmetry springs out in a certain limit of quantum gravity coupled to matter fields after integrating out the gravitational/topological degrees of freedom \cite{a8, chern}. This amounts to having an effective theory in the form of a noncommutative field theory on the $\kappa$-deformed Minkowski space. The second reason is that the DSR models \cite{kgn2002, kgn2003, acmrm2011} are mostly studied within the framework of the $\kappa$-Poincar\'e algebra, where the $\kappa$-deformed Minkowski spacetime provides the arena for studying the particle kinematics.  

$\kappa$-Minkowski space is related to a deformation of the Poincar\'{e} group as a Hopf algebra, called the $\kappa$-Poincar\'{e}-Hopf algebra, which coacts
on $\kappa$-Minkowski space.  The physical implications of $\kappa$-deformations were extensively studied and some of the main results of pursuing this line of research are, e.g., the construction of quantum field theories \cite{lnr1992}-\cite{mstw11}, electrodynamics \cite{h}-\cite{jonke}, NC quantum mechanics \cite{kvantna, kvantna1}, considerations of quantum gravity effects \cite{m2006}-\cite{btz} and the modification of particle statistics \cite{kappaSt}-\cite{Gumesa} on $\kappa$-Minkowski space.

In formulating field theories on NC spaces, differential calculus plays an essential role.
 There is a bicovariant differential calculus \cite{w} on $\kappa$-Poincar\'{e}-Hopf algebra  which in general has an extra cotangent direction (calculus has one dimension more than the classical case) for the time-like deformations \cite{sitarz}. This was generalized to $n$-dimensions in \cite{Gonera}. In \cite{Majid-2} it was discussed that by gauging this extra one form, one can introduce gravity in the model, and in \cite{mercati} this framework is adapted to formulate field theories. Besides that, the physical interpretation of this extra one-form with no classical analogue remains unclear and one is further motivated to construct a NC differential calculus of classical dimension.

In \cite{klry2009, byk2009}, Poincar\'{e} algebra was extended with the dilatation operator to the so called Poincar\'{e}-Weyl algebra, which enabled  the construction of four-dimensional differential algebra on the $\kappa$-Minkowski space. From the action of exterior derivative, in \cite{mkj2009, mkj2011} differential algebras of classical dimension were also constructed.

In \cite{mskj2012} the authors have constructed two families ($\mathcal{D}_{1}$ and $\mathcal{D}_{2}$) of differential algebras of classical dimension. When introducing Lorentz symmetry, the main novelty was that the realization for the Lorentz generators contains Grassmann-type variables. As a consequence, generators of the Lorentz algebra act covariantly on one-forms,
without the need to introduce an extra cotangent direction (extra one-form). The action is also covariant if restricted to the $\kappa$-Minkowski space. However, one loses Lorentz covariance when considering forms of order higher than one.

Bicovariant differential calculus on $\kappa$-Minkowski space-time has been analyzed in \cite{jms2013}. The $\kappa$-deformed coordinates and forms (and also super-Heisenberg algebra), using extended twist, have been constructed.   However, this construction is compatible with $\kappa$-deformed $\mathfrak{igl}(4)$-Hopf algebra.  

In this letter we construct unified graded differential algebra generated by NC coordinates $\hat{x}_{\mu}$, Lorentz generators $M_{\mu\nu}$ and anticommuting NC one-forms $\hat{\xi}_{\mu}$. We demand that $M_{\mu\nu}$ and $\hat{x}_{\mu}$ generate a Lie algebra (compatible with $\kappa$-Poincar\'e-Hopf algebra), $\hat{\xi}_{\mu}$ transform vector-like under Lorentz transformations, and that the algebra between NC coordinates $\hat{x}_{\mu}$ and NC one-forms $\hat{\xi}_{\mu}$ is closed in one-forms (i.e. the differential calculus is of classical dimension). We show that for deformations $a^2\neq 0$ (i.e. time- and space-like) the super-Jacobi identities containing $M_{\mu\nu}$,  $\hat{x}_{\mu}$ and $\hat{\xi}_{\mu}$ are not satisfied. When we enlarge this algebra by introducing a new generator $\hat{\d}$, interpreted as exterior derivative, we find a new unique algebra that satisfies all super-Jacobi identities. This new graded differential algebra is universal, i.e. it is valid for all type of deformations $a_{\mu}$ (time-like, space-like and light-like). When $a_{\mu}=(a_{0},\vec{0})$, the obtained algebra corresponds to the differential algebra in \cite{sitarz} if the extra form $\phi$ is identified with our exterior derivative $\hat{\d}$. Different realizations of our algebra in terms of super-Heisenberg algebra are presented, and it is shown that the volume form $vol$ is undeformed. For light-like deformations $a^2=0$ we get 4-D bicovariant calculus (compatible with \cite{kosinski}), present the corresponding Hopf algebra structure (which is $\kappa$-Poincar\'e-Hopf algebra), and construct the corresponding twist operator. This twist in $a^2=0$ case can be written in a new covariant way and is expressed in terms of Poincar\'e generators only. It is interesting to note that in the case $a^2\neq 0$, this twist leads to $\kappa$-Snyder space.
 
\section{Construction of the unified graded differential algebra}
Let NC coordinates $\hat{x}_{\mu}$ satisfy $\kappa$-Minkowski spacetime algebra, 
\begin{equation}\label{kappa}
[\hat{x}_{\mu},\hat{x}_{\nu}]=i(a_{\mu}\hat{x}_{\nu}-a_{\nu}\hat{x}_{\mu}),
\end{equation}
generators $M_{\mu\nu}$ satisfy Lorentz algebra
\begin{equation}\label{lorentz}
[M_{\mu\nu},M_{\lambda\rho}]=\eta_{\nu\lambda}M_{\mu\rho}-\eta_{\mu\lambda}M_{\nu\rho}-\eta_{\nu\rho}M_{\mu\lambda}+\eta_{\mu\rho}M_{\nu\lambda},
\end{equation}
and basic NC one-forms $\hat{\xi}_{\mu}$ satisfy
\begin{equation}\label{xi}
\left\{\hat{\xi}_{\mu},\hat{\xi}_{\nu}\right\}=0,
\end{equation}
where $a_{\mu}$ are constant deformation parameters, $\eta_{\mu\nu}=\text{diag}(-1,1,1,1)$, $M_{\mu\nu}$ are antisymmetric ($M_{\mu\nu}=-M_{\nu\mu}$) and the Greek indices $\mu,\nu,\rho,...=\left\{0,1,2,3\right\}$. In order to generalize Lie superalgebra generated by $\left\{\hat{x}_{\mu},M_{\mu\nu},\hat{\xi}_{\mu}\right\}$ which should be compatible with $\kappa$-Poincar\'{e}-Hopf algebra \cite{mskj2007} i.e.
\begin{equation}\label{Mx}
[M_{\mu\nu},\hat{x}_{\lambda}]=\eta_{\nu\lambda}\hat{x}_{\mu}-\eta_{\mu\lambda}\hat{x}_{\nu}-ia_{\mu}M_{\nu\lambda}+ia_{\nu}M_{\mu\lambda}
\end{equation}
we impose the Lorentz covariance condition (see \cite{mskj2012} and \cite{jms2013})
\begin{equation}\label{Mxi}
[M_{\mu\nu},\hat{\xi}_{\lambda}]=\eta_{\nu\lambda}\hat{\xi}_{\mu}-\eta_{\mu\lambda}\hat{\xi}_{\nu}
\end{equation}
and the requirement that the commutator $[\hat{\xi}_{\mu},\hat{x}_{\nu}]$ is closed in one-forms $\hat{\xi}_{\lambda}$
\begin{equation}\label{xix}
[\hat{\xi}_{\mu},\hat{x}_{\nu}]=iK_{\mu\nu}^{\ \ \lambda}\hat{\xi}_{\lambda}
\end{equation}
where $K_{\mu\nu}^{\ \ \lambda}\in\mathbb{R}$ and $\hat{x}^{\dagger}_{\mu}=\hat{x}_{\mu}$, $M^{\dagger}_{\mu\nu}=-M_{\mu\nu}$, and $\hat{\xi}^{\dagger}_{\mu}=\hat{\xi}_{\mu}$.
In \cite{mskj2007} it was proven that relations \eqref{kappa}, \eqref{lorentz} and \eqref{Mx} define a unique Lie algebra generated by $\left\{\hat{x}_{\mu},M_{\mu\nu}\right\}$ that  leads to 
$\kappa$-Poincar\'{e} Hopf algebra (after introducing commutative momenta $\hat{p}_{\mu}$). The relations \eqref{lorentz}, \eqref{xi} and \eqref{Mxi} define undeformed Lie superalgebra generated by $\left\{\hat{\xi}_{\mu},M_{\mu\nu}\right\}$. In \cite{mskj2012} two families 
of differential algebras, $\mathcal{D}_{1}$ and $\mathcal{D}_{2}$, generated by  $\left\{\hat{\xi}_{\mu},\hat{x}_{\mu}\right\}$ were constructed. In \cite{arinprogres} it is shown that for $a_{\mu}=(a_{0},\vec{0})$, there exist three family of algebras $\mathcal{D}_{1}$, $\mathcal{D}_{2}$ and $\mathcal{D}_{3}$ which are the only differential algebras compatible with super-Jacobi identities, which exhausts all possible choices of $K_{\mu\nu}^{\ \ \lambda}$. However, we point out that there does not exist Lie superalgebra
generated by $\left\{\hat{x}_{\mu},M_{\mu\nu},\hat{\xi}_{\nu}\right\}$ and satisfying \eqref{kappa}-\eqref{xix} for deformations $a_{\alpha}a^{\alpha}\neq 0$. Namely, mixed Jacobi relations containing $M_{\mu\nu}$, $\hat{x}_{\lambda}$ and $\hat{\xi}_{\rho}$ cannot be satisfied for any choice of $K_{\mu\nu}^{\ \ \lambda}$ corresponding to families $\mathcal{D}_{1}$, $\mathcal{D}_{2}$ and $\mathcal{D}_{3}$ . More precisely, if we postulate $\kappa$-Minkowski space \eqref{kappa}, Lorentz algebra \eqref{lorentz}, then demand that $\hat{x}_{\mu}$ and $M_{\mu\nu}$ form a Lie algebra \eqref{Mx},  that one-forms are closed \eqref{xix} and  transform vector-like under Lorentz transformations (Lorentz condition for bicovariant calculus) \eqref{Mxi}, then the Jacobi identity for $M_{\mu\nu}$, $\hat{x}_{\alpha}$ and $\hat{\xi}_{\beta}$ is not satisfied (after we explicitly use the expression for  $K_{\mu\nu}^{\ \ \lambda}$ corresponding to  algebras $\mathcal{D}_{1,2,3}$ and in general $a_{\alpha}a^{\alpha}\neq 0$). This means that for $a_{\alpha}a^{\alpha}\neq 0$ we cannot have algebra that is closed in one-forms and at the same time compatible with $\kappa$-Poincare algebra i.e. with Lorentz covariance condition \eqref{Mx} and \eqref{Mxi}. However, we can satisfy all super-Jacobi identities for generators $\left\{\hat{x}_{\mu},M_{\mu\nu},\hat{\xi}_{\nu}\right\}$ defined by eqs. \eqref{kappa}-\eqref{xix}  if we do not insist\footnote{ However, if we insist that the algebra is closed in one-forms, then the whole algebra is compatible with $\kappa$-$\mathfrak{igl}(4)$ in general. That is we have
\begin{equation*}\begin{split}
&[M_{\mu\nu}, \hat{x}_{\lambda}]=\eta_{\nu\lambda}\hat{x}_{\mu}-\eta_{\mu\lambda}\hat{x}_{\nu}\\
&\ \ \ \ \ \ \ \ \ \ \ \ \ +\eta_{\nu\lambda}K^{\alpha}_{\ \beta\mu}L_{\alpha}^{\ \beta}-\eta_{\mu\lambda}K^{\alpha}_{\ \beta\nu}L_{\alpha}^{\ \beta}-K^{\alpha}_{\ \nu\lambda}L_{\alpha\mu}+K^{\alpha}_{\ \mu\lambda}L_{\alpha\nu}-K_{\nu\beta\lambda}L_{\mu}^{\ \beta}+K_{\mu\beta\lambda}L_{\nu}^{\ \beta}
\end{split}\end{equation*}
where $L_{\mu\nu}$ is a generator of $\mathfrak{gl}$(4) algebra for which we have 
\begin{equation*}
[L_{\mu\nu},\hat{x}_{\lambda}]=\eta_{\nu\lambda}\hat{x}_{\mu}+\eta_{\nu\lambda}K^{\alpha}_{\ \beta\mu}L_{\alpha}^{\ \beta}-K^{\alpha}_{\ \nu\lambda}L_{\alpha\mu}-K_{\nu\beta\lambda}L_{\mu}^{\ \beta}
\end{equation*}} that the algebra is closed in one forms $\hat{\xi}_{\mu}$.  One possible way to construct such graded algebra is to include commuting momenta $\hat{p}_{\mu}$ that is compatible with relations \eqref{kappa}-\eqref{xix}. Then the commutator between one-forms $\hat{\xi}_{\mu}$ and NC coordinates $\hat{x}_{\mu}$ is no longer closed in one forms, i.e. tensor $K_{\mu\nu}^{\ \ \lambda}$ depends on $\hat{p}_{\mu}$, that is $K_{\mu\nu}^{\ \ \lambda}\equiv K_{\mu\nu}^{\ \ \lambda}(\hat{p})$.
In general, there is infinitely many such graded algebras. For example,  such algebras are defined by  eqs.\eqref{kappa}-\eqref{xix} and momenta $\hat{p}_{\mu}=P_{\mu}$ in natural realization (or classical basis) \cite{kgn2002, mskj2007, mssg2008,  bp2010}
\begin{equation}\begin{split}\label{primjer}
&[P_{\mu},P_{\nu}]=0, \\
&[P_{\mu},\hat{x}_{\nu}]=-i\left(\eta_{\mu\nu}Z^{-1}-a_{\mu}P_{\nu}\right), \quad [P_{\mu}, \hat{\xi}_{\nu}]=0,\\
&[M_{\mu\nu},P_{\lambda}]=\eta_{\nu\lambda}P_{\mu}-\eta_{\mu\lambda}P_{\nu}
\end{split}\end{equation}
where $Z^{-1}=a\cdot P+\sqrt{1+a^2P^2}$, with one explicit example
 \begin{equation}\label{primjer1}
K_{\mu\nu}^{\ \ \lambda}(P)=\eta_{\mu\nu}a^{\lambda}+\delta^{\lambda}_{\nu}a_{\mu}+a^2 P_{\mu}\delta^{\lambda}_{\nu}+\frac{a^2 P_{\nu}\delta^{\lambda}_{\mu}}{\sqrt{1+a^2P^2}}
\end{equation}
 (for another example see eqs. (46-48) in \cite{jms2013}).

There is another way of constructing Lie superalgebra, by adding the new generator $\hat{\d}$, interpreting it as NC exterior derivative, satisfying the following relations
\begin{equation}\begin{split}\label{d}
&\ \ \hat{\d}^2=0, \quad [\hat{\d},\hat{x}_{\mu}]=\hat{\xi}_{\mu},\\
&[M_{\mu\nu},\hat{\d}]=0, \quad \left\{\hat{\d},\hat{\xi}_{\mu}\right\}=0.
\end{split}\end{equation}
Now, we write the most general ansatz for the commutator between one-forms $\hat{\xi}_{\mu}$ and NC coordinates $\hat{x}_{\mu}$ that is compatible with the Lorentz algebra, i.e. $[\hat{\xi}_{\mu},\hat{x}_{\nu}]=i\left[C_{1}a_{\mu}\hat{\xi}_{\nu}+C_{2}a_{\nu}\hat{\xi}_{\mu}+\eta_{\mu\nu}\left(C_{3}(a\cdot\hat{\xi})+C_{4}a^2\hat{\d}\right)\right]$, where $C_{1,...,4}\in\mathbb{R}$, and impose the super-Jacobi identities.
Then the solution for parameters $C_{1,...,4}$ is unique and  the commutation relation is given by
\begin{equation}\label{xixd}
[\hat{\xi}_{\mu},\hat{x}_{\nu}]=ia_{\mu}\hat{\xi}_{\nu}-\eta_{\mu\nu}\left(i(a\cdot\hat{\xi})+a^2\hat{\d}\right).
\end{equation}
The Lie superalgebra $\left\{\hat{x}_{\mu},M_{\mu\nu}, \hat{\xi}_{\mu},\hat{\d}\right\}$  defined by \eqref{kappa}-\eqref{Mxi}, \eqref{d} and \eqref{xixd} is the unique superalgebra compatible with $\kappa$-Poincare Hopf algebra\footnote{For details on how to extract the coalgebra, that is, coproducts from commutation relations, see \cite{domagoj}.} (after including momenta, for example like $P_{\mu}$ in \eqref{primjer}). In \cite{jms2013} it is pointed out that starting with $\kappa$-Poincar\'{e} Hopf algebra and the action\footnote{For the definition and properties of the action $\blacktriangleright$ see \cite{jms2013}.} $M_{\mu\nu}\blacktriangleright \hat{x}_{\lambda}=\eta_{\nu\lambda}\hat{x}_{\mu}-\eta_{\mu\lambda}\hat{x}_{\nu}$, $M_{\mu\nu}\blacktriangleright \hat{\xi}_{\lambda}=\eta_{\nu\lambda}\hat{\xi}_{\mu}-\eta_{\mu\lambda}\hat{\xi}_{\nu}$ and $M_{\mu\nu}\blacktriangleright \hat{\d}=0$ one can obtain the commutations relations \eqref{Mx}, \eqref{Mxi} and $[M_{\mu\nu},\hat{\d}]=0$.
The commutation relations \eqref{xixd} contain all commutation relations in \cite{sitarz}  (see eq. (54) in \cite{sitarz} or eq. (22) in \cite{jms2013}) with identifying $\phi=-\hat{\d}$ and $a_{\mu}=(a_{0},\vec{0})$.

\section{Realizations in terms of super-Heisenberg algebra}
The simplest realization of $\left\{\hat{x}_{\mu},M_{\mu\nu}, \hat{\xi}_{\mu},\hat{\d}\right\}$ in terms of super-Heisenberg algebra is given by
\begin{equation}\begin{split}\label{nat}
&\hat{x}_{\mu}=X_{\mu}Z^{-1}+i(a\cdot X)D_{\mu}-i\xi_{\mu}(a\cdot q)+i(a\cdot \xi)q_{\mu}+\frac{a^2(\xi\cdot D)q_{\mu}}{\sqrt{1-a^2D^2}}\\
&M_{\mu\nu}=X_{\mu}D_{\nu}-X_{\nu}D_{\mu}+\xi_{\mu}q_{\nu}-\xi_{\nu}q_{\mu}\\
&\hat{\xi}_{\mu}=\xi_{\mu}, \quad \hat{\d}=\xi_{\alpha}\frac{D^{\alpha}}{\sqrt{1-a^2D^2}}
\end{split}\end{equation}
where we have used the natural realization for  $\left\{X,D\right\}$, i.e. \eqref{primjer} with $P_{\mu}=-iD_{\mu}$ (see also \cite{kgn2002, mskj2007, mssg2008, bp2010}) and $\left\{X,D,\xi,q\right\}$ satisfy
\begin{equation}\begin{split}\label{X}
&[X_{\mu},X_{\nu}]=[X_{\mu},\xi_{\nu}]=[X_{\mu},q_{\nu}]=0,\\
&[D_{\mu},D_{\nu}]=[D_{\mu},\xi_{\nu}]=[D_{\mu},q_{\nu}]=0,\\
&\left\{\xi_{\mu},\xi_{\nu}\right\}=0, \quad \left\{q_{\mu},q_{\nu}\right\}=0,\\
&[D_{\mu},X_{\nu}]=\eta_{\mu\nu}, \quad \left\{q_{\mu},\xi_{\nu}\right\}=\eta_{\mu\nu}.
\end{split}\end{equation}
Using similarity transformations $\mathcal{S}$ (automorphisms of super-Heisenberg algebra)
\begin{equation}\begin{split}
&x^{\prime}_{\mu}=\mathcal{S}x_{\mu}\mathcal{S}^{-1}, \quad \partial^{\prime}_{\mu}=\mathcal{S}\partial_{\mu}\mathcal{S}^{-1},\\
&\xi^{\prime}_{\mu}=\mathcal{S}\xi_{\mu}\mathcal{S}^{-1}, \quad q^{\prime}_{\mu}=\mathcal{S}q_{\mu}\mathcal{S}^{-1},
\end{split}\end{equation}
where $\left\{x,\partial,\xi,q\right\}$ and $\left\{x^{\prime},\partial^{\prime},\xi^{\prime},q^{\prime}\right\}$ are different realizations of super-Heisenberg algebra satisfying relations like \eqref{X}, we can obtain infinite family of realizations\footnote{For discussion on realizations and similarity transformations for Heisenberg algebra see \cite{twist2013} and for super-Heisenberg algebra \cite{jms2013}.} for generators  $\left\{\hat{x}_{\mu},M_{\mu\nu}, \hat{\xi}_{\mu},\hat{\d}\right\}$.
For instance, note that if we use $X_{\mu}^{\prime}\equiv\left(X_{\mu}-a^2(X\cdot D)D_{\mu}\right)\sqrt{1-a^2D^2}$ and $D_{\mu}^{\prime}\equiv\frac{D_{\mu}}{\sqrt{1-a^2D^2}}$ we have
\begin{equation}\begin{split}\label{crt}
&\hat{x}_{\mu}=X^{\prime }_{\mu}-i\left(X^{\prime}_{\mu}(a\cdot D^{\prime})+\xi_{\mu}(a\cdot q)\right)+\left(i(a\cdot X^{\prime})+a^2(X^{\prime}\cdot D^{\prime})\right)D^{\prime}_{\mu}+\left(i(a\cdot \xi)+a^2(\xi\cdot D^{\prime})\right)q_{\mu}\\
&M_{\mu\nu}=X^{\prime}_{\mu}D^{\prime}_{\nu}-X^{\prime}_{\nu}D^{\prime}_{\mu}+\xi_{\mu}q_{\nu}-\xi_{\nu}q_{\mu}\\
&\hat{\xi}_{\mu}=\xi_{\mu}, \quad \hat{\d}=\xi^{\alpha}D^{\prime}_{\alpha}
\end{split}\end{equation}
It is easy to see that $X_{\mu}^{\prime}$ and $D_{\mu}^{\prime}$ generate Heisenberg algebra.
On the other hand if we use the bicrossproduct basis \cite{majid} (right ordering \cite{mssg2008}) we have
\begin{equation}\begin{split}\label{bicros}
&\hat{x}_{i}=x_{i}-i\xi_{i}(a\cdot q)+i(a\cdot \xi)q_{i}+\frac{a^2(\xi\cdot D)q_{i}}{\sqrt{1-a^2D^2}}\\
&\hat{x}_{0}=x_{0}+ia_{0}x_{k}\partial_{k}-i\xi_{0}(a\cdot q)+i(a\cdot \xi)q_{0}+\frac{a^2(\xi\cdot D)q_{0}}{\sqrt{1-a^2D^2}}\\
&M_{ij}=x_{i}\partial_{j}-x_{j}\partial_{i}+\xi_{i}q_{j}-\xi_{j}q_{i}\\
&M_{i0}=x_{i}\left(\frac{1-Z}{ia_{0}}+\frac{ia_{0}}{2}\square Z\right)-\left(x_{0}+ia_{0}x_{k}\partial_{k}\right)\partial_{i}+\xi_{i}q_{0}-\xi_{0}q_{i}\\
&\hat{\xi}_{\mu}=\xi_{\mu}, \quad \hat{\d}=\xi_{\alpha}\frac{D^{\alpha}}{\sqrt{1-a^2D^2}},
\end{split}\end{equation}
where
\begin{equation}\begin{split}
&D_{i}=\partial_{i}Z^{-1}, \quad D_{0}=\frac{Z^{-1}-1}{ia_{0}}+\frac{ia_{0}}{2}\Box,\\
&\Box=\partial_{k}^{2}Z^{-1}+\frac{4}{a^{2}_{0}}\text{sinh}^2\left(\frac{ia\cdot\partial}{2}\right), \quad Z=\text{e}^{ia\cdot\partial}
\end{split}\end{equation}
In all three examples above, we have used a special type of similarity transformations which transform only the generators of Heisenberg algebra and leave the Grassmann part unchanged. Alternatively,  we  can fix Heisenberg algebra (for example $\left\{X_{\mu}, D_{\nu}\right\}$), and apply similarity transformation to Grassmann part $\left\{\xi_{\mu},q_{\nu}\right\}$ such that  
\begin{equation}\label{dopcenito}
\hat{\xi}_{\mu}=\xi^{\prime}_{\mu}\Sigma(D^2), \quad \hat{\d}=\xi^{\prime}_{\alpha}\Sigma(D^2)\frac{D^{\alpha}}{\sqrt{1-a^2D^2}}
\end{equation}
where $\Sigma(D^2)$ is an arbitrary function of $D^2$ satisfying $\underset{a\to 0}{\lim}\left(\Sigma(D^2)\right)=1$. The above relations could all be written in terms of physical momentum, that is $P_{\mu}=-iD_{\mu}$, $P^{\prime}_{\mu}=-iD^{\prime}_{\mu}$ and $p_{\mu}=-i\partial_{\mu}$.

Note that if we would have NC coordinates without the Grassmann part, then the corresponding exterior derivative would not be compatible with the Lorentz covariance conditions \eqref{Mx}, \eqref{Mxi} and \eqref{d} (see also \cite{mskj2012}, \cite{jms2013}).

\section{Outlook and discussion}
Our results can be summarized as:
\begin{enumerate}
\item By embedding the whole algebra $\left\{\hat{x}_{\mu},M_{\mu\nu}, \hat{\xi}_{\mu},\hat{\d}\right\}$ into the super-Heisenberg algebra, we can express   the exterior derivative $\hat{\d}$ in terms of one forms $\xi_{\alpha}$ and functions of momenta (see \eqref{nat}, \eqref{crt}, \eqref{bicros} and \eqref{dopcenito}). This way we can interpret relation \eqref{xixd} as \eqref{xix} but with $K_{\mu\nu}^{\ \ \lambda}(P)=a_{\mu}\delta^{\lambda}_{\nu}-\eta_{\mu\nu}\left(a^{\lambda}+a^2\frac{P^{\lambda}}{\sqrt{1+a^2P^2}}\right)$, which illustrates that eqs. \eqref{Mx}, \eqref{Mxi} and super-Jacobi identities imply that the $K_{\mu\nu}^{\ \ \lambda}$ cannot just be a constant, i.e. the algebra is not closed in one-forms $\hat{\xi}_{\mu}$. In general there are infinitely many such algebras (for another explicit example see \eqref{primjer1}).
\item The additional one-form $\phi$ in Sitarz's approach \cite{sitarz} is exterior derivative $\hat{\d}$ in our approach, which is a direct consequence of the definition of exterior derivative \eqref{d} and  leads to $\hat{\d}\blacktriangleright 1=0$. 
Furthermore $vol=\hat{\xi}_{0}\hat{\xi}_{1}\hat{\xi}_{2}\hat{\xi}_{3}$ is invariant under the action of Lorentz generators $M_{\mu\nu}$ and $\hat{\d}\ vol=vol\ \hat{\d}=0$ in our approach, and in \cite{mercati} following Sitarz's approach $vol^5=\phi\ \ vol=vol\ \phi\neq 0$.
\item In our approach we have defined commutation relations $[M_{\mu\nu},\hat{x}_{\lambda}]$, $[M_{\mu\nu},\hat{\xi}_{\lambda}]$ and $[M_{\mu\nu},\hat{\d}]$, whereas such relations do not exist in \cite{sitarz}\footnote{The general procedure of how to obtain the commutation relations from coproduct using action $\blacktriangleright$ is elaborated in \cite{pla2013}, and for this particular case see \cite{jms2013}.}.
\item We have obtained realizations of $\hat{x}_{\mu}$ ,$M_{\mu\nu}$, $\hat{\xi}_{\mu}$ and $\hat{\d}$ in terms of super-Heisenberg algebra. We point out that we can choose $vol=\hat{\xi}_{0}\hat{\xi}_{1}\hat{\xi}_{2}\hat{\xi}_{3}=\xi_{0}\xi_{1}\xi_{2}\xi_{3}$, but the realizations of $\hat{x}_{\mu}$ and $M_{\mu\nu}$ always contain Grassmann part (in \cite{mskj2012} $\hat{x}$ has not contained Grassmann part).

\end{enumerate}

In this letter we have presented a new unique Lie superalgebra generated by $\left\{\hat{x}_{\mu},M_{\mu\nu}, \hat{\xi}_{\mu},\hat{\d}\right\}$ and  defined by 
\begin{equation}\begin{split}\label{lie}
&[\hat{x}_{\mu},\hat{x}_{\nu}]=i(a_{\mu}\hat{x}_{\nu}-a_{\nu}\hat{x}_{\mu}),\quad  \left\{\hat{\xi}_{\mu},\hat{\xi}_{\nu}\right\}=0,\\
&[M_{\mu\nu},M_{\lambda\rho}]=\eta_{\nu\lambda}M_{\mu\rho}-\eta_{\mu\lambda}M_{\nu\rho}-\eta_{\nu\rho}M_{\mu\lambda}+\eta_{\mu\rho}M_{\nu\lambda},\\
&[M_{\mu\nu},\hat{x}_{\lambda}]=\eta_{\nu\lambda}\hat{x}_{\mu}-\eta_{\mu\lambda}\hat{x}_{\nu}-ia_{\mu}M_{\nu\lambda}+ia_{\nu}M_{\mu\lambda}\\
&[M_{\mu\nu},\hat{\xi}_{\lambda}]=\eta_{\nu\lambda}\hat{\xi}_{\mu}-\eta_{\mu\lambda}\hat{\xi}_{\nu}, \quad [M_{\mu\nu},\hat{\d}]=0,\\
&[\hat{\d},\hat{x}_{\mu}]=\hat{\xi}_{\mu}, \quad \left\{\hat{\d},\hat{\xi}_{\mu}\right\}=0,\\
&[\hat{\xi}_{\mu},\hat{x}_{\nu}]=ia_{\mu}\hat{\xi}_{\nu}-\eta_{\mu\nu}\left(i(a\cdot\hat{\xi})+a^2\hat{\d}\right)
\end{split}\end{equation}
Algebra \eqref{lie} satisfies all super-Jacobi identities and, with commutative momenta (e.g. \eqref{primjer}), represents a unique differential algebra satisfying bicovariant calculus compatible with $\kappa$-Poincar\'{e}-Hopf algebra. We emphasize  that \eqref{lie} is valid for general deformation vector $a_{\mu}$, i.e. time-like, space-like and light-like.

Note that, by acting with the Lie superalgebra \eqref{lie}  on $\blacktriangleright 1$, we define a reduced algebra generated by $\left\{\hat{\xi}_{\mu}, \hat{x}_{\nu}\right\}$ identical with  \cite{beggs2013} (see equation (6) in \cite{beggs2013}).  This reduced algebra is nonassociative\footnote{ The only algebras that are closed in one-forms and associative (i.e. satisfy super-Jacobi identities) are $\mathcal{D}_{1}$, $\mathcal{D}_{2}$ and $\mathcal{D}_{3}$ (for time-like deformations $a_{\mu}$). In \cite{beggs2013}, equation (6),  an algebra closed in one-forms is presented, but it is not included in the families $\mathcal{D}_{1}$, $\mathcal{D}_{2}$ and $\mathcal{D}_{3}$, hence it is nonassociative (i.e. it does not satisfy super-Jacobi identities). } for $a^2\neq 0$. The construction in \cite{beggs2013} which leads to this reduced algebra is based on the requirement that the Minkowski metric is a central element of this  graded algebra ( in the first order in the deformation parameter). This approach  leads to 
formulation of Schr\"{o}dinger equation with modified mass, modified Klein-Gordon equation, and deformed electrodynamics.

In \cite{beggs} there is a construction of general noncommutative metric for a special kind of differential algebra, i.e. $[\hat{\xi}_{\mu},\hat{x}_{\nu}]=ia_{\mu}\hat{\xi}_{\nu}$, which belongs to the family  $\mathcal{D}_{2}$ (see \cite{mskj2012}). The construction is based on the requirement that the metric is a central element in this differential algebra. We plan to generalize this construction for families $\mathcal{D}_{1}$, $\mathcal{D}_{2}$ and $\mathcal{D}_{3}$ \cite{arinprogres}, as well as for \eqref{lie}.

Finally, we present the unique differential algebra compatible with bicovariant calculus and $\kappa$-Poincar\'{e}-Hopf algebra, which is closed in basic one forms $\hat{\xi}_{\mu}$. Namely, if the deformation vector $a_{\mu}$ is light-like (e.g. $a_{\alpha}a^{\alpha}\equiv a^2=0$), then the commutation relations \eqref{xixd} are closed in one-forms $\hat{\xi}_{\mu}$ i.e.
\begin{equation}\label{xixa}
[\hat{\xi}_{\mu},\hat{x}_{\nu}]=ia_{\mu}\hat{\xi}_{\nu}-i\eta_{\mu\nu}(a\cdot\hat{\xi})
\end{equation}
and satisfy all super-Jacobi identities. So, the graded differential algebra \eqref{lie} with $a^2=0$ is associative. Furthermore, the realizations, eqs. \eqref{nat} and \eqref{crt} are identical and reduce to
\begin{equation}\begin{split}
&\hat{x}_{\mu}=X_{\mu}+ia^{\alpha}M_{\alpha\mu}\\
&M_{\mu\nu}=X_{\mu}D_{\nu}-X_{\nu}D_{\mu}+\xi_{\mu}q_{\nu}-\xi_{\nu}q_{\mu}\\
&\hat{\xi}_{\mu}=\xi_{\mu}, \quad \hat{\d}=\xi_{\alpha}D^{\alpha}
\end{split}\end{equation}
where we have used $Z^{-1}=1-i(a\cdot D)$ and $\square=D^2$. It seems that the graded differential algebra written in the new covariant form \eqref{lie}, with $a^2=0$, i.e. closed in one-forms $\hat{\xi}_{\mu}$, was overlooked in the literature so far, but it is compatible with the $4D$ bicovariant calculus presented in \cite{kosinski}.  We point out that the 4D graded differential algebra over $\kappa$-Minkowski space which possesses bicovariant calculus invariant under the $\kappa$-Poincar\'e-Hopf algebra exists only for light-like deformations ($a^2=0$).

The generators $M_{\mu\nu}$ and $P_{\mu}$ satisfy undeformed Poincar\'{e} algebra, see eqs. \eqref{lorentz} and \eqref{primjer}. The corresponding $\kappa$-deformed Poincar\'{e}-Hopf algebra can be written in the unified covariant way \cite{Govindarajan-2, mskj2007, mssg2008, domagoj} and the coproduct $\Delta$ is given by
\begin{equation}\begin{split}\label{coproduct}
\Delta P_{\mu}&=P_{\mu}\otimes Z^{-1}+1\otimes P_{\mu}-a_{\mu}\left(P_{\alpha}-\frac{a_{\alpha}}{2}\square\right)Z\otimes P^{\alpha},\\
\Delta M_{\mu\nu}&=M_{\mu\nu}\otimes 1+1\otimes M_{\mu\nu}-a_{\mu}\left(P^{\alpha}-\frac{a^{\alpha}}{2}\square\right)Z\otimes M_{\alpha\nu}+a_{\nu}\left(P^{\alpha}-\frac{a^{\alpha}}{2}\square\right)Z\otimes M_{\alpha\mu},
\end{split}\end{equation} 
as well as the antipode $S$ and counit $\epsilon$
\begin{equation}\begin{split}\label{antipod}
&S(P_{\mu})=\left[-P_{\mu}-a_{\mu}\left(P_{\alpha}-\frac{a_{\alpha}}{2}\square\right)P^{\alpha}\right]Z,\\
&S(M_{\mu\nu})=-M_{\mu\nu} -a_{\mu}\left(P^{\alpha}-\frac{a^{\alpha}}{2}\square\right) M_{\alpha\nu}+a_{\nu}\left(P^{\alpha}-\frac{a^{\alpha}}{2}\square\right) M_{\alpha\mu},\\
&\epsilon(P_{\mu})=\epsilon(M_{\mu\nu})=0.
\end{split}\end{equation}
The above Hopf algebra structure unifies all three types of deformations  $a_{\mu}$, i.e. time-like ($a^2<0$), space-like ($a^2>0$) and light-like ($a^2=0$). The twist operator $\mathcal{F}$ for the time-like case was constructed in the Hopf algebroid approach in \cite{twist2013}. In the case $a^2\neq 0$ twist operator $\mathcal{F}$ cannot be expressed in terms of Poincar\'{e} generators only. However, if $a^2=0$ the twist operator $\mathcal{F}$ can be written in a new covariant way in terms of Poincar\'{e} generators only 
\begin{equation}\label{twist}
\mathcal{F}=\text{exp}\left\{a^{\alpha}P^{\beta}\frac{\text{ln}(1+a\cdot P)}{a\cdot P}\otimes M_{\alpha\beta}\right\},
\end{equation}
and satisfies cocycle condition. This twist operator generates $\kappa$-deformed Poincar\'{e}-Hopf algebra, defined by eqs.\eqref{coproduct}, \eqref{antipod}, with $Z^{-1}=1+a\cdot P$, $\square=-P^2$, coproduct $\Delta g=\mathcal{F}\Delta_{0}g\mathcal{F}^{-1}$,  antipode $S(g)=\chi \ S_{0}(g)\ \chi^{-1}$ (where $\chi= m\left[\left(1 \otimes S_{0}\right)\mathcal{F}\right]$), and counit $\epsilon(g)=0$, where $\Delta_{0}g=g\otimes 1+1\otimes g$ is primitive coproduct and $S_{0}(g)=-g$ undeformed antipode  for $g\in \left\{M_{\mu\nu}, P_{\mu}\right\}$. In the light-cone basis the twist element  is  written in a non-covariant form in \cite{universal} (see eq. (41) in \cite{universal}).
The full graded differential algebra for $a^2=0$ with the action $\blacktriangleright$ leads to the super-Hopf algebroid structure, which will be elaborated separately.

It is interesting to note that the  twist operator in \eqref{twist} for $a^2\neq 0$ gives $\hat{x}_{\mu}=X_{\mu}+ia^{\alpha}M_{\alpha\mu}$ and $\hat{\xi}_{\mu}=\xi_{\mu}$ (satisfying \eqref{xixa}) and leads to $\kappa$-Snyder space \cite{kappasnyder} defined by
\begin{equation}\label{snyder}
[\hat{x}_{\mu},\hat{x}_{\nu}]=i(a_{\mu}\hat{x}_{\nu}-a_{\nu}\hat{x}_{\mu})+a^2M_{\mu\nu}, \quad [P_{\mu},\hat{x}_{\nu}]=-i\eta_{\mu\nu}(1+a\cdot P)+ia_{\mu}P_{\nu}.
\end{equation}
The twist operator \eqref{twist} for $a^2\neq 0$ does not satisfy cocycle condition and leads to nonassociative star product.
 This is a new result and the details of this construction  and corresponding Hopf algebra/algebroid structure will be presented elsewhere.

Note that all above results (especially \eqref{lie}-\eqref{snyder}) are valid in arbitrary dimension $n\geq 2$, and for all non-Euclidean signatures of the metric $\eta_{\mu\nu}$. 

In the forthcoming work we plan to  use our approach in developing the notions of Lie derivative, inner product, cyclic property for integration and  Hodge star , generalizing the results in \cite{mercati}. Also, developing the notion of  NC Dirac operator \cite{dirac}, star product realization and extended twist operator  is crucial for applying the above approach to NCQFT \cite{klry09, Govindarajan-2, ms11, mstw11}, NC gravity \cite{a4, a5, a6, a9, a10, beggs} and relative locality \cite{acfkgs2011, mpsg}.

\bigskip

\noindent{\bf Acknowledgment}\\
The authors would like to thank Kumar S. Gupta, Anna Pachol and Zoran Skoda for useful comments.  R. S. would like to thank the Ruder Boskovic Institute for the kind hospitality during
the course of this work. \\

\end{document}